\documentclass{ws-ijmpc}

\usepackage{amssymb}
\usepackage{comment}
\usepackage{eurosym}   %per il simbolo dell'euro
\usepackage{mathrsfs}  %per gli integrali funzionali
\DeclareRobustCommand{\Eqref}[1]{Eq.~(\ref{#1})}

\usepackage{booktabs} %per tabelle fancy
\usepackage{dcolumn} 
\newcolumntype{d}[1]{D{.}{.}{#1}}
\usepackage{multirow}

\usepackage{todo}     %per mettere note (utile solo nel draft)
\usepackage{soul}     %per mettere testo barrato o evidenziato (utile solo nel draft)

\usepackage[hyphens]{url} %per usare links nelle referenze o altrove

\usepackage{courier}

\usepackage{float}
\usepackage{listings} %per mettere dei pezzi di codice belli da vedere
\usepackage{lineno}

\usepackage[binary-units=true]{siunitx}
\usepackage{algorithmic}
\usepackage{algorithm}
\usepackage[normalem]{ulem}
\usepackage[english]{babel}
\usepackage[caption=false]{subfig}

\def\eg{\emph{e.g.} }
\def\ie{\emph{i.e.} }

\begin{document}

%\linenumbers % switch on line numbering for the whole article

\markboth{C. Bonati \& E. Calore  \& M. D'Elia \& M. Mesiti \& 
  F. Negro \& F. Sanfilippo \& S.F. Schifano \& G. Silvi \& R. Tripiccione}
{Portable multi-node LQCD Monte Carlo simulations using OpenACC}

\catchline{}{}{}{}{}
%%%%%%%%%%%%%%%%%%%%%%%%%%%%%%%%%%%%%%%%%%%%%%%%%%%%%%%%%%%%%%%%%%%%

\title{Portable multi-node LQCD Monte Carlo simulations using OpenACC}

%%%%%%%%%%%%%%%%%%%%%%%%%%%%%%%%%%%%%%%%%%%%%%%%%%%%%%%%%%%%%%%%%%%%%%%%

\author{Claudio Bonati}
\address{Universit\`a di Pisa and INFN Sezione di Pisa,\\ 
Largo Pontecorvo 3, I-56127 Pisa, Italy\\
claudio.bonati@df.unipi.it
}
\author{Enrico Calore}
\address{Universit\`a degli Studi di Ferrara and INFN Sezione di Ferrara,\\
Via Saragat 1, I-44122 Ferrara, Italy\\
enrico.calore@fe.infn.it}
\author{Massimo D'Elia}
\address{Universit\`a di Pisa and INFN Sezione di Pisa,\\ 
Largo Pontecorvo 3, I-56127 Pisa, Italy\\
massimo.delia@unipi.it
}
\author{Michele Mesiti}
\address{Academy of advanced computing, Swansea University, \\
Singleton Park, Swansea SA2 8PP, UK\\
michele.mesiti@swansea.ac.uk
}
\author{Francesco Negro}
\address{INFN Sezione di Pisa,\\ 
Largo Pontecorvo 3, I-56127 Pisa, Italy\\
fnegro@pi.infn.it
}
\author{Francesco Sanfilippo}
\address{INFN Sezione di Roma3,\\ 
Via della Vasca Navale 84, I-00146 Roma, Italy\\
sanfilippo@roma3.infn.it
}
\author{Sebastiano Fabio Schifano}
\address{Universit\`a degli Studi di Ferrara and INFN Sezione di Ferrara,\\ 
Via Saragat 1, I-44122 Ferrara, Italy\\
schifano@fe.infn.it
}
\author{Giorgio Silvi}
\address{J{\"u}lich Supercomputing Centre, 
Forschungszentrum J{\"u}lich,\\
Wilhelm-Johnen-Stra{\ss}e, 52428 J{\"u}lich, Germany \\ 
g.silvi@fz-juelich.de
}
\author{Raffaele Tripiccione}
\address{Universit\`a degli Studi di Ferrara and INFN Sezione di Ferrara,\\
Via Saragat 1, I-44122 Ferrara, Italy\\
tripiccione@fe.infn.it
}

%%%%%%%%%%%%%%%%%%%%%%%%%%%%%%%%%%%%%%%%%%%%%%%%%%%%%%%%%%%%%%%%%%%%%%%%

\maketitle

\begin{history}
\received{Day Month Year}
\revised{Day Month Year}
\end{history}

%\newpage %% DA RIMUOVERE

%%%%%%%%%%%%%%%%%%%%%%%%%%%%%%%%%%%%%%%%%%%%%%%%%%%%%%%%%%%%%%%%%%%%%%%%

\begin{abstract}

This paper describes a state-of-the-art parallel Lattice QCD Monte Carlo 
code for staggered fermions, purposely designed to be portable across 
different computer architectures, including GPUs and commodity CPUs.
Portability is achieved using the OpenACC parallel programming model, 
used to develop a code that can be compiled for several processor 
architectures. 
The paper focuses on parallelization on multiple computing nodes using 
OpenACC to manage parallelism within the node, and OpenMPI to manage 
parallelism among the nodes. 
We first discuss the available strategies to be adopted to maximize performances, 
we then describe selected relevant details of the code, and finally measure 
the level of performance and scaling-performance that we are able to achieve.
The work focuses mainly on GPUs, which offer a significantly high level 
of performances for this application, but also compares with results measured 
on other processors.

% Max 4 / 5 keywords: 

\end{abstract}

\ccode{PACS Nos.: 
07.05.Bx %Computer systems: hardware, operating systems, computer languages, and utilities. 
12.38.Gc %Lattice QCD calculations
}

%%%%%%%%%%%%%%%%%%%%%%%%%%%%%%%%%%%%%%%%%%%%%%%%%%%%%%%%%%%%%%%%%%%%%%%%

\section{Introduction}

Monte Carlo simulations play a key role in the
study of several aspects of Quantum Chromodynamics (QCD), the quantum
field theory that describes the strong interaction in the standard model
of particle physics.
Lattice QCD is a computational scheme based on importance
sampling Monte Carlo simulations used to study QCD in
the non-perturbative regime, \ie
when the theory is strongly interacting and perturbation theory cannot be
applied.  This approach has been extremely effective in obtaining first
principles calculations of the hadron masses\cite{Bazavov:2009bb,Fodor:2012gf}
and of the thermodynamical properties of the quark-gluon
plasma\cite{Ding:2016qdj}, while technical difficulties are still encountered when a
large baryon chemical potential is present\cite{Aarts:2015tyj}. Despite the
continuous efforts in the development of more and more efficient algorithms\cite{KennedyLec}, 
Lattice QCD simulations still require a tremendous amount of computing resources.

Lattice QCD simulations belong to the class of HPC grand challenge applications, 
with physics results strongly limited by available computational
resources\cite{Bernard:2002pd,bilardi}. For this reason in the mid '80s 
the increasing request for computational resources triggered the development 
of massively parallel supercomputers~\cite{nicola,qcdoc,apecise,bgl,qpace08,bgq} 
specifically designed and optimized to match the computing requirements of 
Lattice QCD algorithms. 
This approach lost its effectiveness as general purpose supercomputers started to 
be available on the market; interestingly enough, the architectures of 
these commercial machines were still similar to the ones of their LQCD-optimized 
predecessors, with low frequency and low power CPUs in each node and toroidal 
interconnection between the nodes.
Nowadays we are in the middle of a new change of paradigm in the field of High 
Performance Computing (HPC), in which the typical node-to-node communication
harness is all-to-all, and ``slim'' computing nodes are being replaced by
``fat'' ones, based on multi- and many-core CPUs or Graphic Processing Units (GPUs). 

While the most recent CPU and GPU architectures seem to follow a converging 
evolutionary path, with an always increasing level of vectorization, different
programming languages are still used to develop applications for CPUs and GPUs.
This poses significant portability issues, that need to be seriously addressed, as 
there is still no clear consensus on the computing architectures -- CPUs or GPUs -- 
that will be mostly adopted in the near future.

The OpenACC\cite{openacc} standard is a solution designed to address portability 
issues across several computing architectures, using a programming model similar 
to OpenMP\cite{openmp}, but specifically designed to allow applications to run on accelerators. 
It abstracts code functionalities to a descriptive level, leaving architecture-specific 
implementations to the compiler. 
Using this approach, the same source code may run on all processors, GPUs and also CPUs, 
supported by the compiler, achieving an easy and good level of code portability.
OpenACC is becoming increasingly popular among several scientific communities 
for coding many lattice-based applications to run mainly on GPU accelerators, 
including Lattice Boltzmann Methods\cite{blair15,jiri,ccpe16}, and more recently 
also Lattice QCD~\cite{gupta-lqcd-openacc,lqcdoacc17}.

The change from ``slim'' to ``fat'' nodes, however, does not affect only the 
portability of the code, but also its parallel structure. In fact: 
i) the traditional strategy to minimize the surface over volume ratio of the tiles 
is no more a priori the optimal approach to get the best parallel efficiency, 
ii) different levels of parallelization must be exploited, 
ii) and data organization plays an increasing relevant role for computing 
performances. 

In our previous work\cite{lqcdoacc17} we have described an OpenACC 
implementation of a state-of-the-art Monte Carlo Lattice QCD application, 
derived from an earlier version coded using CUDA\cite{incardona}, and able 
to run only on single-accelerator systems, including GPUs (NVIDIA and AMD), 
and also CPUs.
In this paper we extend our OpenACC code to run also on accelerators-based 
parallel computing machines, discussing in detail how we have structured the 
code, the strategies that have guided our design choices, and presenting 
several performance results on different computing architectures.
We focus mainly on GPU-based clusters since for this kind of applications 
they offer a level of performances an order of magnitude higher than standard 
high-end commodity CPUs. 
We describe how we distribute the workload of the application among the GPUs, 
the strategies used to overlap communications and computation, 
and analyze computing efficiency and strong scalability on several nodes. 
As in the previous work, we use the PGI compiler able to target all versions of  
NVIDIA GPUs as well as commodity CPUs, offering code portability and the 
possibility to measure, compare and analyze also performance figures of clusters 
based on recent multi-core Intel Xeon CPUs.

The structure of this paper is as follows: in Sec.~\ref{sec:alg} we 
briefly recall the global structure of Lattice QCD simulation algorithm,
mainly focusing on those aspects that are relevant for parallelization; in
Sec.~\ref{sec:parallelization} we analyze several parallelization strategies 
to distribute the data-domain over several computing-nodes and the tools 
available to exchange data, and in Sec.~\ref{sec:parallel-implementation} 
we describe the details of our implementation.
Finally, in Sec.~\ref{sec:perf} we analyze the parallel efficiency and 
scaling figures of the code, and in Sec.~\ref{sec:conclusions} we draw 
our conclusions.

%%%%%%%%%%%%%%%%%%%%%%%%%%%%%%%%%%%%%%%%%%%%%%%%%%%%%%%%%%%%%%%%%%%%%%%%

\section{Algorithms for Lattice QCD}\label{sec:alg}

In this section we summarize the basic algorithmic ingredients of a
typical Lattice QCD simulation that are needed to understand the parallel
structure of our implementation.  We will use the same notation
of a previous paper describing a single device implementation\cite{lqcdoacc17}, to 
which we refer for more details and reference to the original 
literature\cite{RotheBook,DeGrandDeTarBook}.

In a Lattice QCD simulation the space-time continuum is approximated by a
lattice of spacing $a$ and extents $aN_t, aN_x, aN_y, aN_z$. The fundamental
variables are the gauge fields $U_{\mu}(x)$, which are $3 \times 3$ unitary complex
matrices associated to the links $(x,\mu)$
of the lattice ($x$ is a site of lattice and $\mu$ a direction), the momenta
$H_{\mu}(x)$ conjugated to the gauge fields ($3 \times 3$ complex Hermitian
and traceless matrices) and the pseudofermions $\phi(x)$,
associated to the site $x$ of the lattice. These variables have to be sampled 
according to the probability distribution (for the case of a single staggered fermion)
\begin{equation}\label{eq:prob}
P(U, H, \phi)\propto\exp\left(-\frac{1}{2} H^2 -S_g[U]-\phi^{\dag} M[U^{(k)}]^{-1/4}\phi\right)\ ,
\end{equation}
where $H^2$ stands for the sum over the whole lattice  
of the traces of the squared momenta, while 
the scalar function $S_g(U)$ (the gauge part of the action) is a sum of
traces of path-ordered products along closed circuits of the gauge variables
$U_{\mu}(x)$.  In our code we used for $S_g$ the so-called tree-level Symanzik
improved discretization\cite{Weisz:1982zw,Curci:1983an}, in which only $1\times
1$ and $1\times 2$ rectangular paths (with all possible orientations) enter
the action. 

The last term in \Eqref{eq:prob} is the fermion part of the action and
$M[U^{(k)}]$ is the Dirac matrix: in the staggered discretization this matrix
connects nearest neighbor sites of the lattice and the hopping term between
the sites $x$ and $x+\mu$ is proportional to the $k-$times stout
smeared\cite{Morningstar:2003gk} gauge field $U^{(k)}_{\mu}(x)$ (with the
convention $U_{\mu}^{(0)}(x)\equiv U_{\mu}(x)$, the case $k=0$ corresponds to
the simple staggered fermions). Since $M[U^{(k)}]$ connects only nearest
neighbor sites it is convenient to use an even-odd
preconditioning\cite{DeGrandDeTarBook,DeGrand:1990dk}; it can be
shown that it is sufficient to have pseudofermions associated only to the even
sites of the lattice. In the following we will denote by $D_{oe}$ and $D_{eo}$ the two
out-of-diagonal blocks of $M[U^{(k)}]$ (note, for future reference, that $D_{oe}^{\dag}=-D_{eo}$)

We use the Rational Hybrid Monte Carlo
algorithm\cite{Clark:2004cp,Clark:2006fx,Clark:2006wp} (RHMC) to sample the
probability distribution in Eq.~\eqref{eq:prob}, using a Markov chain Monte
Carlo approach: the fractional power of $M(U)$ is approximated to machine
precision by a rational function of $M(U)$ and the update is performed by a
combination of Molecular Dynamics (MD) evolution of the gauge fields and
accept/reject steps, like in the ordinary Hybrid Monte Carlo
update\cite{Duane:1987de,KennedyLec}. Pseudofermions enter quadratically in the
action, so they are generated by an heatbath step at the beginning of the MD
evolution and remain constant along the MD trajectory.

Most of the simulation time is spent in the computation of the force acting on
the gauge field (needed for the MD evolution) and in the
evaluation of the action at the end of the trajectory, needed for the
accept/reject step. These high level operations map at intermediate level
to the following two operations: products of $U_{\mu}(x)$ matrices along some
simple paths and solutions of linear equations of the form:
\begin{equation}\label{eq:shift_eo}
(m^2\, I-D_{eo}D_{oe}+\sigma^{(i)})\varphi^{(i)}=b \ ,
\quad i\in\{1,\ldots,r\}\ , 
\end{equation}
where $m$ is the fermion mass, $r$ is the order of the rational approximation
used in the RHMC and $\sigma^{(i)}$ are the positions of the poles of the
rational approximation. Since the linear operators acting on the left-hand side
of \Eqref{eq:shift_eo} are positive definite, these equations can be
conveniently solved by using the shift (also known as multi-mass) form of the
Conjugate Gradient\cite{Jegerlehner:1996pm,Simoncini}, whose elementary
building blocks are vector linear algebra (basically scalar products and sums)
and the application of the matrices $D_{oe}$ and $D_{eo}$ to a vector. 

Also for the multi-node implementation, as for the single-node, several
algorithmic improvement can be implemented on top of the basic scheme described
so far, however these improvements typically do not require any additional
effort in the parallelization. Features that are implemented but not described
here for this reason are multi-step\cite{Sexton:1992nu,Urbach:2005ji} and
improved integrators\cite{Omelyan02,Omelyan03,Takaishi:2005tz}, the use of
multiple pseudofermions\cite{Clark:2006fx} and of different values of the
stopping residuals and of the rational approximation orders in different parts
of the RHMC\cite{Clark:2006wp}.

%%%%%%%%%%%%%%%%%%%%%%%%%%%%%%%%%%%%%%%%%%%%%%%%%%%%%%%%%%%%%%%%%%%%%%%%

\section{Parallelization and data exchange on regular lattices}
\label{sec:parallelization}

In this section we first analyze design options for the development of a multi-process
parallel version of our LQCD code\cite{lqcdoacc17}, and then we give an overview of the 
available tools for data communications in parallel systems.

\subsection{Strategies for parallelization}
\label{sec:strategies}

Designing a parallel multi-process LQCD code is in principle straightforward\cite{bilardi}. 
One splits the physical lattice in regular tiles of the same size, and 
maps them onto a cluster of processing nodes. Doing that, the processing load is balanced 
among the processing elements, and communication patterns among the computing nodes are regular, 
predictable and only involve (logically) nearest-neighbor processes.
However, if processes reside on different computing nodes, node-to-node communication  
introduces overheads that may seriously hamper the scaling behavior of the code, 
as the number of processing nodes increases. 

Consider a lattice of $N$ points in $D$ dimensions (\ie with linear size $L = N^{1/D}$,   
$D = 4$ in our case), and parallelize it on $d \le D$ dimensions, dividing the lattice 
in regular sub-lattice tiles and mapping them onto $N_p$ processors.
As information exchange is roughly proportional to the surface of each computational 
domain, while computing grows as the domain volume, one should in principle minimize 
the surface-over-volume ratio $(S/V) \simeq d ~ N_p^{1/d}$. One is then led to the 
conclusion that the largest possible choice for $d$ ($d = D$, $d = 4$ in our case) 
should have the best asymptotic scaling behavior.
This depends on the apparently obvious assumption that node-to-node communication 
bandwidth does not depend on the size of data-messages, and on  
the way the physical lattice is tiled. 
However, this is not necessarily the case for currently available large cluster systems, 
for many reasons: 
\begin{itemize}
\item communication of data buffers, corresponding to physical surfaces not
stored in contiguous memory locations, implies a gather-and-scatter overhead
that may seriously reduce sustained communication bandwidth;
\item as one increases $d$, the size of each communication chunk becomes
smaller; however, since communication functions have large startup latencies,
this reduces effective sustained bandwidth\cite{parco16};
\item current available multi- and many-core processors have large memories and high 
computing-power, making the computation more coarse-grained compared to previous machines, such as
several generations of Blue-Gene systems.
Near-peak sustained performance on these processors implies substantial streaming computation, thus 
each processor need to handle a large enough data-domain, limiting the number of computing nodes 
that can be used for many lattice-domain sizes of interest from the physics point of view;
\item tiling the lattice domain on many dimensions implies a significantly more
complex code structure, that may hamper further optimization
steps\cite{hpcs15}.
\end{itemize}
For these reasons, in this work we have decided to tile our lattice in just one dimension ($d = 1$), 
leaving all three remaining dimensions fully contained within each processing node. 
However we have taken care to allow an arbitrary mapping among the code coordinates ($0, 1, 2, 3$) 
and the physical ones ($t,x,y,z$), so one can select which physical coordinate should be tiled onto 
the processors; this is obviously relevant when using asymmetric lattices.

For further analysis, we now consider the amount of data items that must be
transferred across node boundaries for the most compute intensive operations
that we have described in the previous section.  Consider first the evaluation
of the Dirac operator; since the matrices $D_{oe}$ and $D_{eo}$ connect only
neighboring sites (and pseudofermions are constant along the MD evolution), in
the solution of \Eqref{eq:shift_eo} we need only to communicate (after each
application of $D_{eo}$ or $D_{oe}$) the values of the pseudofermions in a
slice of thickness $1$ along the boundaries.  Since the pseudofermions field is
represented by $3$ \texttt{double complex} numbers for each even site, we must
transfer 48 bytes for unit of surface in both directions.  Some communication
is obviously needed also in the computation of the scalar products, but this is
negligible with respect to the previous one.

A larger amount of data transfer is required for the gauge fields: the function
$S_g$ uses the products of gauge fields along $1\times 1$ and $1\times 2$
rectangles; thus in the computation of the (gauge part of the) force acting on
$U_{\mu}(x)$ we need the values of the gauge field at sites that are up to $2$
lattice spacing away from $x$.  As a consequence, for the computation of the so
called ``staples'' (that are in the LQCD context the equivalent of the stencils
for partial differential equations), we need to communicate the values of the
gauge fields in a slice of thickness $2$, for a total amount of 768 bytes
(using gauge link compression) for unit of surface. 

The other relevant strategy for scaling performance is the implementation of overlap 
between computations and node-to-node communications.
Consider for instance the evaluation of the Dirac operator; the physical points
sitting on a contact-surface between two processing elements have a data
dependency with the adjoining physical points, sitting on the
logically-neighbor processor.  It is customary to organize \emph{halo}-regions,
containing updated copies of the data corresponding to those points; the
obvious strategy is that one: i) applies first the Dirac operator to the points
belonging to the contact-surface and then, ii) applies the Dirac operator to
all other bulk lattice points and \emph{at the same time} transfers the freshly
computed data values to the corresponding halos on the neighbor processors.

Summing up, in our 1-d tiling approach on $N_p$ processors, for a given lattice size, communication 
time $T_c$ is roughly constant in time, while the processing time $T_p$ decreases as $1/N_p$; we 
then expect for the total processing time $T_T \simeq max ~\left[ T_c, T_p(N_p)\right]$, that is 
(nearly) perfect scaling as long as $T_c \le T_p$ followed by a regime in which adding 
processors yields almost no performance improvement. 

%%%%%%%%%%%%%%%%%%%%%%%%%%%%%%%%%%%%%%%%%%%%%%%%%%%%%%%%%%%%%%%%%%%%%%%%
\subsection{Tools for data exchange}
\label{sec:data-exchange}

Large GPU clusters are widely heterogeneous computing systems, with compute nodes 
hosting one or more CPU processors, each acting as host for a variable number of GPUs 
directly connected to their host through the PCIe bus interface, together with the 
network interface, such as Infiniband.
 
The complexity of this structure implies that what, at the application level,  
is a plain GPU-to-GPU communication, may involve different hardware routes, 
different communication protocols, and correspondingly different performances 
both in terms of latency and bandwidth. 
A large development effort has been put in place in recent years to make communication 
programming relatively transparent to hardware details and at the same time provide a reasonable
level of efficiency. We use several such developments in our code and briefly describe them here.
Current implementations of MPI, like OpenMPI indeed, support several features to allow for easy 
and efficient communications among GPUs. The following are included in our code: 
\begin{itemize}
\item
CUDA-aware MPI\cite{cuda-aware_mpi} allows to specify buffers allocated on the 
GPU memory as arguments of the MPI operations, making codes terser and more readable.
\item
For GPUs attached to the same host, CUDA-IPC moves data directly across GPUs
without staging on CPU memory, making communication faster\cite{parco16}.
\item
GPUs attached to different CPUs of the same node communicate through 
CPU-memory staging, pipelining all communication steps to shorten communications 
latency.
\item
For GPUs belonging to different nodes, GPUDirect RDMA\cite{rdma} moves short data packets 
from the GPU to the network interface without any involvement of the host CPU. 
For longer data packets due to PCIe architectural bottlenecks, RDMA becomes 
less effective\cite{rdma}. 
In this case, GPUDirect simplifies the operation by sharing a common staging 
region between the GPU and the network interface.
\end{itemize}

%%%%%%%%%%%%%%%%%%%%%%%%%%%%%%%%%%%%%%%%%%%%%%%%%%%%%%%%%%%%%%%%%%%%%%%%

\section{Parallel Implementation}
\label{sec:parallel-implementation}

Our code uses plain C99 language, the standard MPI library and OpenACC.  The
MPI library is used to perform the first coarse tiling of the lattice, with
tiles of equal size along one direction as described in
Sec.~\ref{sec:parallelization}.  Different processes operate on different
sub-lattices, with each process typically associated to one processor unit (\eg
a CPU or a GPU) and communications between neighboring processes being managed by the
MPI library. OpenACC directives are used to take care of the parallelization across
the computing elements of each single processing unit (\eg a CPU or a GPU).

OpenACC\cite{openacc} is a directive based language abstracting parallel
programming to a descriptive level, relieving programmers from specifying how
codes should be mapped onto the target architecture. It is similar to OpenMP
and was introduced to manage parallelism on accelerators, such as GPUs,
although it is designed to be architecture agnostic\cite{Wienke2014812}, and
the same code can be compiled and run also on standard CPU processors.
For more detail on the OpenACC implementation see our previous work\cite{lqcdoacc17}.

The RHMC algorithm conceptually consists of two different units, namely
molecular dynamics (MD) and the Metropolis test. In the Metropolis test the
value of the action (\ie the exponent in Eq.\ref{eq:prob}) has to be computed
before and after the MD evolution and the most compute-intensive part of this
unit is the solution of a  linear systems of the form in
Eq.~\eqref{eq:shift_eo}. As noted in Sec.~\ref{sec:alg} this basically amount
to repeated applications of the linear operators $D_{eo}$ and $D_{oe}$, which
connect nearest neighbors lattice sizes through the $U^{(k)}$ link matrices.

In the MD step the main task is to update the SU(3) link matrices $U^{(0)}$ in
Eq.\ref{eq:prob}, which is done by solving a set of first-order differential
equations which involve the gauge links and their conjugate momenta. In the
computation of the force driving the MD evolution two terms are present: one
coming from $\phi^{\dag} M[U^{(k)}]^{-1/4}\phi$ (the \emph{fermionic} term),
and the other coming from $S_g(U)$ (the \emph{gauge} term).  As far as the
fermionic term is concerned, the most compute-intensive step is the already
discussed multi-shift Conjugate Gradient solver for the Dirac operator. In our
lattice discretization of the theory (\ie the tree-level improved Symanzik
gauge action) the evaluation of the gauge term of the force requires the computation
of products of $U^{(0)}$ link matrices along the perimeter of $1\times 1$
plaquettes and $2\times 1$ rectangles. Once the force acting on each link is computed,
the MD algorithm proceeds in an embarrassingly parallel way until the next
computation of the force is required.

Another part of the algorithm which is not embarrassingly parallel is the one
related to the so-called stouting procedure\cite{Morningstar:2003gk}, which
enters the algorithm in two places. The first place is the computation of the
$k$-stouted links $U^{(k)}$ (which are used in the Dirac operator) starting
from the original links $U^{(0)}$ which enter the gauge term.  The second place
is the computation of the force in the MD evolution, since the fundamental
variables to evolve are the $U^{(0)}$ links, but the fermionic part of the
action depends on $U^{(k)}$ (see \cite{Morningstar:2003gk} for the procedure to be used).
For both these computations data relative to a 1-site thick halos have to be communicated.
Since however the computing time spent in the stouting procedure is roughly one
order of magnitude smaller than the time spent in the pure gauge molecular
dynamics evolution and in the Dirac operator, the impact on the global
performance of any optimization of the communication pattern used in this step
would be minimal.

In the following we will focus on the communication-related aspects of the
implementation (for a more detailed description of the algorithm see our
previous paper\cite{lqcdoacc17} or the standard references \cite{RotheBook,DeGrandDeTarBook}).

%%%%%%%%%%%%%%%%%%%%%%%%%%%%%%%%%%%%%%%%%%%%%%%%%%%%%%%%%%%%%%%%%%%%%%%%

\subsection{Data structures, domain partitioning and data compression}
\label{sec:data-struct}

\begin{figure}[ht]
\centering
  \subfloat[The vec3\_soa data structure]{
     \includegraphics[width=0.8\textwidth]{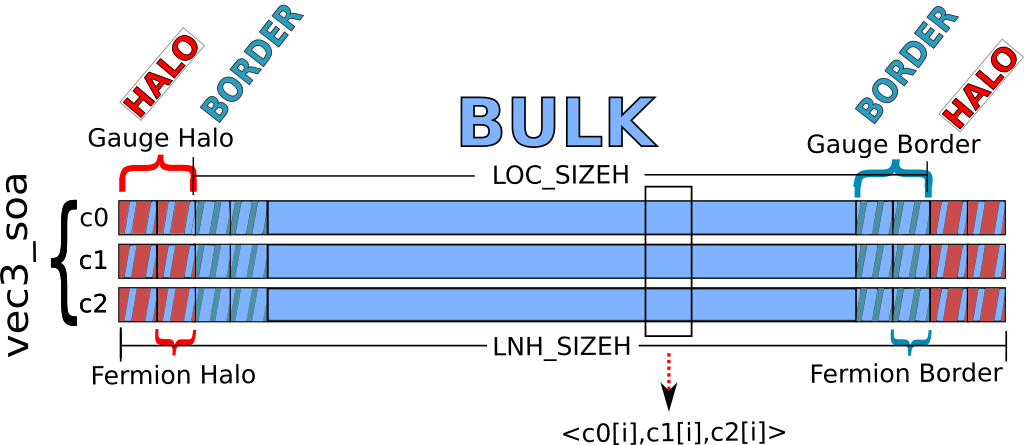}
     \label{fig:vec3_soa}
  } \\
  \subfloat[The su3\_soa data structure]{
     \includegraphics[width=0.8\textwidth]{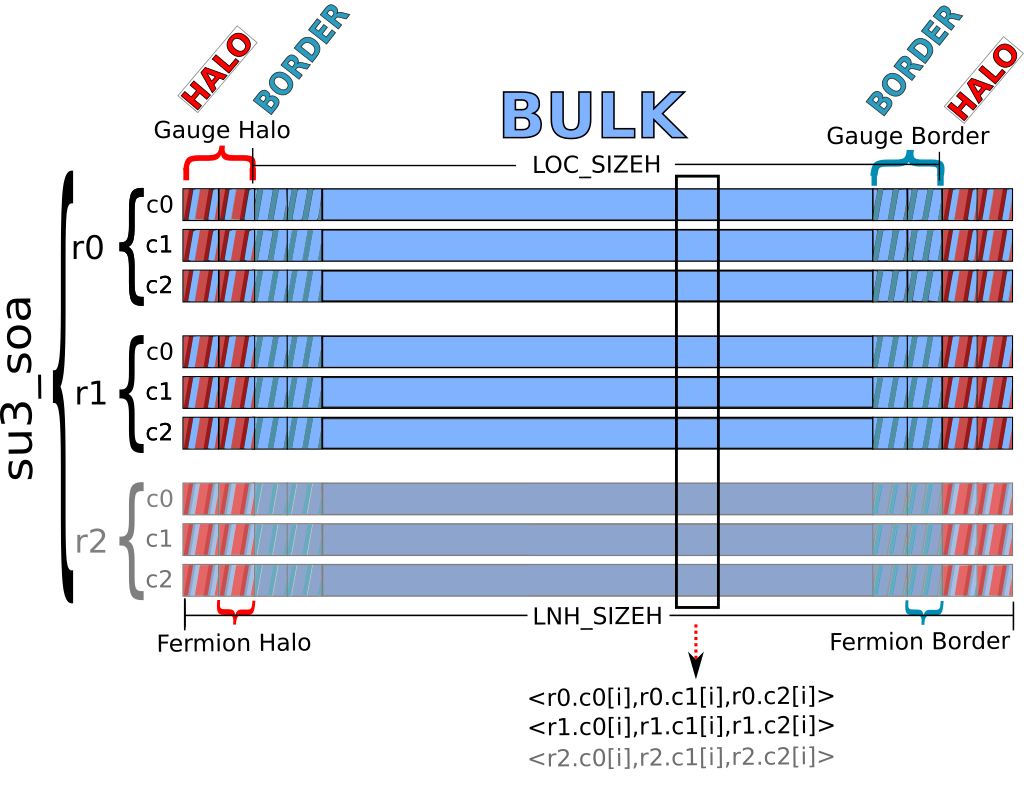}
     \label{fig:su3_soa}
  } \\
  
\caption{Graphical view of how data structures corresponding to 
the vec3\_soa and the su3\_soa data-types are stored in memory. 
This layout allow to exploit instruction vectorization of the code.}
\label{fig:datastructs}
\end{figure}

Thanks to the 1d-tiling approach we adopt, there is no need for gather-scatter
operations since data to be moved between processes are already at contiguous
memory locations. A graphic view  of the data structures used in the code is
shown in Fig.\ref{fig:datastructs}. \\ We store the pseudofermions fields is
the {\sf vec3\_soa } structure (see listing \ref{lst:soadeclarations} and
\figurename~\ref{fig:datastructs}). It consist of 3 arrays of double (or float)
C99 complex numbers, arranged in a  Structure of Arrays (SoA)
layout\cite{se4hpcs15,lqcdoacc17}.  Each array has {\sf LNH\_SIZEH
$=n_{0,loc}n_{1,loc}n_{2,loc}(n_{2,loc}+2h)/2$} elements, where $n_{i,loc}$ are
the sizes of the local lattice  and $h$ is the ``largest'' needed halo size,
which is 1 for the Dirac operator, and 1 or 2 for the gauge part depending on
the choice of $S_g$. \\ The data pertaining to the two halo regions for the
current process are stored in the first and last section of the array (red in
\figurename~\ref{fig:datastructs}). The local data domain also splits in three
parts: two ``borders'' (blue in \figurename~\ref{fig:datastructs}), that are
the parts of the local domain corresponding to the halos for neighboring
processes, and a ``bulk'' region, which is not involved in communications. \\
We store data related to the SU(3) matrices representing the gauge
configuration using the {\sf su3\_soa} structure, consisting of 3 {\sf
vec3\_soa} structures corresponding to the rows of the matrix\footnote{This
data structure is also used to store temporary results of the computations
which are GL(3) matrices.}.

We use several techniques to reduce the amount of data exchanged with memory
and with neighbor nodes, in an attempt to increase overall performance.
Indeed, all performance critical kernels are strongly memory-bound on currently
available processors (for instance the Dirac operator has an arithmetic
intensity which is $\approx 6$ times lower than the machine
balance\cite{machine-balance} on recent GPUs).

For the gauge variables, we make use of \emph{gauge link compression}: the
third row of an $SU(3)$ matrix can be computed from the first two as $r3 = (r1
\wedge r2)^\ast$ thanks to the orthonormality conditions, so we do not read it
from memory and we do not include it in communications. This technique is
successfully used to alleviate  bandwidth related issues, \eg in the GPU
implementation of the whole algorithm, and in MPI communications.

A different compression technique is used for quark-related variables. Remember
that each quark flavor is related to a U(1) field $u_\mu (x) = \exp (i
\theta_\mu (x))$, by which we handle antiperiodic boundary conditions in the
time-direction, staggered phases, the imaginary chemical potential and
background  (electro-)magnetic fields.  Hence, while applying the Dirac
operator, pseudofermions must be multiplied by $u_\mu (x) U_\mu (x) $  instead
of $U_\mu (x)$.  We reduce data access request by storing in memory
$\theta_\mu (x) $ and recomputing $u_\mu (x) =\cos \theta_\mu (x) + i \sin
\theta_\mu (x) $ on-line.  Even if trigonometric functions are very
compute-intensive (approximately 15(30) floating point operations in
single(double)-precision) this approach leaves the code memory-bound and
increases performance.

\begin{lstlisting}[language=C,label=lst:soadeclarations,belowcaptionskip=1em,
caption={The implementation of our most important data structures.}]

typedef struct vec3_soa_t{
	double complex c0[LNH_SIZEH],c1[LNH_SIZEH],c2[LNH_SIZEH];
} vec3_soa;

typedef struct su3_soa_t{
	vec3_soa r0,r1,r2;
} su3_soa;

\end{lstlisting}

\subsection{The $D_{oe}$ and $D_{eo}$ operators}

The application in the local domain of the $D_{oe}$ and $D_{eo}$ operators
requires halos which are 1-site thick. Both functions are split into 3 pieces,
two of which compute the result on the borders (using also data from the halos)
and one computes the result on the bulk (which needs no data from the halos).
After the computation ends, all halos must be updated: freshly computed data
belonging to the borders must be sent to neighboring processes while new halo
data must be retrieved from them. A schematic of this procedure is shown in
Algorithm \ref{alg:doe}.  The data to be transferred from and to each process
in this case consists of 6 packets of size $8~n_{0,loc} n_{1,loc} n_{2,loc} $
bytes, as seen in \figurename~\ref{fig:datastructs} (a typical size is around
250KB).

\begin{algorithm}
\caption{Parallelization of the $D_{oe}$ operator in our 1D-tiling approach (the $D_{eo}$ case is 
the same).\label{alg:doe} }
\begin{algorithmic}[1]
\STATE $\phi^{(1)}_{o,l} = D_{oe, l} \phi^{(0)}_e $ on the lower border (on stream 1)
\STATE $\phi^{(1)}_{o,u} = D_{oe, u} \phi^{(0)}_e $ on the upper border (on stream 2)
\STATE $\phi^{(1)}_{o,b} = D_{oe, b} \phi^{(0)}_e $ in the bulk (on stream 3)
\STATE Wait on stream 1 and 2
\STATE Start asynchronous communications of fermion halos of $\phi^{(1)}_{o}$, that is 
$\phi^{(1)}_{o,l}$ and $\phi^{(1)}_{o,e}$
\COMMENT{6 send requests, 6 recv requests} 
\STATE Wait on stream 3 and on the 12 requests
\end{algorithmic}
\end{algorithm}

\subsection{The molecular dynamics evolution}

For the pure gauge part of the molecular dynamics evolution, each iteration of the 
algorithm involves three steps: computing the forces from the gauge configuration at
fictitious time $t$, evolving the momenta according to the evaluated forces to time 
$t+\delta t_1$, and finally evolving the gauge configuration to $t+\delta t_2$ (where $\delta t_1$ and 
$\delta t_2$ depend on the algorithm chosen). Notice that this is a constrained system. The 
first step is the only one that involves inter-process data moves: the so-called staples (which
consist of products of all but one links along the perimeter of a rectangle) must be computed 
for $1 \times 1$ and $ 1\times 2$ rectangles\footnote{The $1\times 2$ staples are only needed 
in case the tree-level Symanzik improved action is used.}. For this step an update of the halos
of the gauge configuration is required.
The gauge configuration consists of 8  {\sf su3\_soa} structures depicted in 
\figurename~\ref{fig:datastructs}. The data to be transferred to and from each process consists of 96 
packets, having size $8 n_{0,loc} n_{1,loc} n_{2,loc} $ bytes (as in the case of the $D_{oe}$ 
and $D_{eo}$ operators).
In order to allow for the superposition of communications of halos with computation in the bulk, 
we adopted the procedure described in Algorithm \ref{alg:gauge}: all three steps are completed first on
the borders, communications of the updated gauge configuration halos are started, the three steps
are performed on the bulk, and then we wait for completion of all communication steps before starting the next iteration. 

\begin{algorithm}
\caption{Parallelization of a block of pure-gauge molecular dynamics in our 1D-tiling approach.\label{alg:gauge} }
\begin{algorithmic}[1]
\STATE gauge force on the lower border 
\STATE gauge force on the upper border 
\STATE new momenta on the lower border 
\STATE new momenta on the upper border 
\STATE new configuration on the lower border 
\STATE new configuration on the upper border 
\STATE start asynchronous communications of new gauge configuration halos
\COMMENT{96 send requests, 96 recv requests} 
\STATE force on the bulk
\STATE new momenta on the bulk
\STATE new configuration on the bulk
\STATE Wait on the 192 requests
\end{algorithmic}
\end{algorithm}

%%%%%%%%%%%%%%%%%%%%%%%%%%%%%%%%%%%%%%%%%%%%%%%%%%%%%%%%%%%%%%%%%%%%%%%%

\section{Performance results}
\label{sec:perf}

In this section we initially describe the computing system we have used to run
all the simulations, \ie the COKA cluster, and then we analyze scaling and
performance figures of the MPI-OpenACC application described in the previous
sections. 

In most Lattice QCD simulations one wants to complete a given number of Monte
Carlo trajectories on a lattice of a specific size in the shortest possible
time, so we focus on \textit{Strong Scaling}, that is we analyze the compute
time as a function of the number of compute devices used to solve the same
problem size, as one splits the same lattice in smaller and smaller tiles.  We
study in larger details the two most time consuming phases of the code -- the
Dirac operator and the pure gauge molecular dynamics, describing their scaling
behavior -- but we also show performance results for the full code running on
thermalized configurations with typical state-of-the-art physics parameters.

We analyze in finer details performance -- and performance bottlenecks -- for
GPUs, since these processors have by far higher sustained performance than
other architectures. However, since our code is fully portable to X86 CPUs we
also show some results for them.

\subsection{The COKA cluster}
\label{sec:coka}

%% Cluster COKA
All our tests have been done on the COKA cluster, a GPU-based HPC cluster jointly operated by INFN 
and Universit\`a degli Studi di Ferrara, with a peak performance of $\approx 100$ TFLOPs.

The COKA cluster has 5 computing nodes, each node embedding $2 \times$ Intel Xeon 
E5-2630v3 CPUs and $8 \times$ NVIDA K80 dual-GPU boards. Each board hosts 
$2 \times$ GK210 GPUs, so there are $16$ CUDA devices on each node.
Nodes are interconnected with 56Gb/s FDR InfiniBand links; each node has
$2 \times$ Mellanox MT27500 Family [ConnectX-3] HCA, allowing \textit{multirail 
networking}\cite{multirail-ib} for a doubled inter-node bandwidth.
The two InfiniBand HCAs are connected respectively to the two PCIe root complexes, connected on 
their turn to the two CPU sockets. 
This allows for a symmetric hardware configuration, where each GPU has one local InfiniBand HCA, 
connected to the same PCIe root complex, so data messages do not need to traverse the inter-socket 
communication link (\ie the Intel Quick Path Interconnect in this case).
In all tests we have used the OpenMPI library, version 1.10.7, exploiting its CUDA-aware
MPI capabilities, when running on GPUs.

\subsection{The Dirac Operator}
\label{sec:dirac-scaling}

In this section we measure the \textit{Strong Scaling} behavior and aggregate performance of the
Dirac operator.
We consider two different lattice sizes \ie $32^3 \times 48$ and $32^3 \times 64$,
which are relevant for physics simulations, and easily divisible across various numbers of GPUs.
We split the former lattice on 1, 2, 4, 6, 8 and 12 GPUs, and the latter
lattice on 2, 4, 8 and 16 GPUs.

\begin{figure}[ht]
\centering
\includegraphics[width=0.8\textwidth]{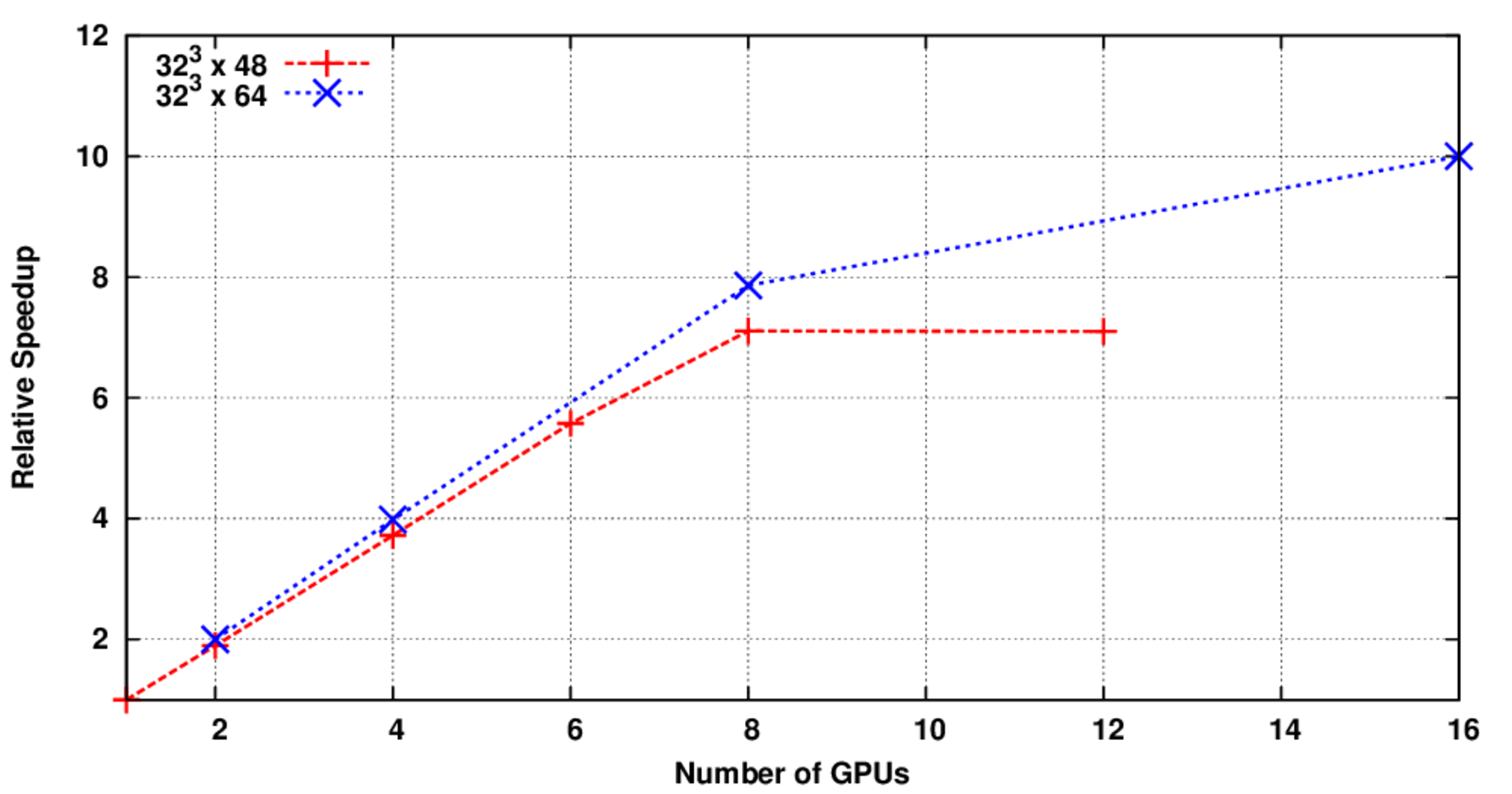}
\caption{Relative Speedup of the Dirac Operator in double precision on two 
lattices of size $32^3 \times 48$ and $32^3 \times 64$, for a growing number of 
GPUs hosted in the same compute node of the COKA Cluster.}
\label{fig:dirac-speedup}
\end{figure}

As discussed in Sec.~\ref{sec:parallelization}, this kernel scales perfectly 
as long as communication time is hidden by computing time over the bulk of the lattice.
This is confirmed in \figurename~\ref{fig:dirac-speedup}, where we can see a perfect 
\textit{Strong Scaling} behavior up to $8$ GPUs for both  lattice sizes.
Further increasing the number of GPUs does not increase the performance anymore
for the $32^3 \times 48$ lattice, thus the speedup reaches a plateau.
On the other hand, for the bigger $32^3 \times 64$ lattice, using more than $8$ GPUs we
can still have a performance increase, although using $16$ GPUs we are far from an optimal 
speedup.

\begin{figure}[ht]
\centering
  \subfloat[One iteration using $8$ GPUs. The time-lines of the kernels executing on two 
  neighboring GPUs are shown, one over the other.]{
     \includegraphics[width=0.95\textwidth]{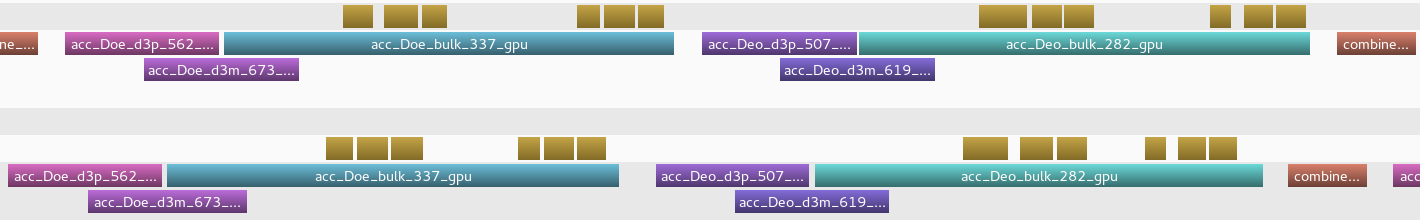}
     \label{fig:dirac-8}
  } \\
  \subfloat[One iteration using $12$ GPUs. The time-lines of the kernels executing on two 
  neighboring GPUs are shown, one over the other.]{
     \includegraphics[width=0.95\textwidth]{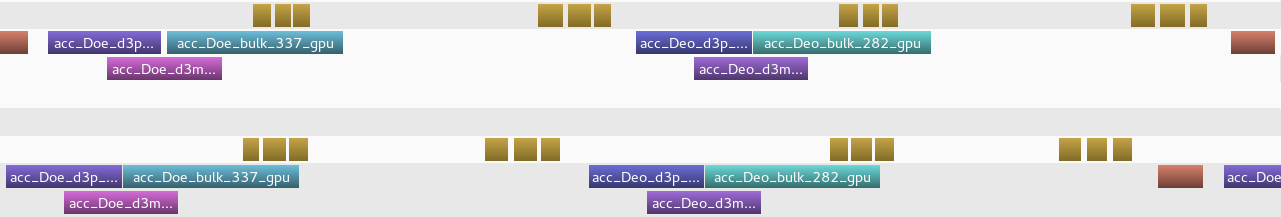}
     \label{fig:dirac-12}
  }
\caption{Time-line of the execution of the Dirac operator on a lattice of $32^3 \times 48$ sites 
using different numbers of GPUs ($8$ in Fig.~\ref{fig:dirac-8} and $12$ in Fig.~\ref{fig:dirac-12}). 
Purple-blue colored: execution of $D_{eo}$ and $D_{oe}$ on the borders of the lattice; 
turquoise colored: execution of $D_{eo}$ and $D_{oe}$ operations on the bulk of the lattice; 
gold colored: communication steps, as seen from the GPUs.
Bulk operations are fully (\ref{fig:dirac-8}) or partially (\ref{fig:dirac-12}) overlapped with 
communication.}
\label{fig:dirac-profile}
\end{figure}

To shed more light on this behavior we use the PGI Profiler to extract traces
of the GPU kernel executions from an actual run of the code.
From \figurename~\ref{fig:dirac-speedup}, we know that for a lattice of $32^3
\times 48$ sites the Dirac operator scales up to $8$ GPUs.  Thus we profiled
two different runs, using the same lattice size, and using respectively $8$ and
$12$ GPUs, looking for execution differences which could explain the scaling
impairment.  Results are shown in \figurename~\ref{fig:dirac-profile}, clearly
showing to which extent the computation phases overlap with communication in
the two different cases.

The Dirac Operator has 7 different kernels, shown in different colors and labeled in 
\figurename~\ref{fig:dirac-8}. The first three blocks on the left, \textit{acc\_Doe\_d3p}, 
\textit{acc\_Doe\_d3m} and \textit{acc\_Doe\_bulk}, refer to the execution of the $D_{oe}$ kernel, 
respectively 
on the borders and on the bulk of the lattice, while \textit{acc\_Deo\_d3p}, 
\textit{acc\_Deo\_d3m} and \textit{acc\_Deo\_bulk}, show $D_{eo}$, again 
on the borders and on the bulk of the lattice.
Eventually, a final kernel, shown in red, is run for each iteration, performing just a 
\textit{zaxpy} operation, corresponding to the final sum of Eq.~\ref{eq:shift_eo}. 

Unlabeled yellow bars, represent MPI communications, as seen from the GPUs
point of view; there are 6 communication steps for each iteration of $D_{eo}$
or $D_{oe}$, as expected.  More interestingly, we can see that in
\figurename~\ref{fig:dirac-8} communications are fully overlapped in time with
the kernels operating on the bulk of the lattice.  On the other hand, in
\figurename~\ref{fig:dirac-12}, the execution time of those same kernels take a
shorter amount of time, as we use a higher number of GPUs, so tiles are
smaller.  In the latter case, since communication time is approximately
constant w.r.t. the tile size, the data transfer step is not completely hidden
behind computation time on the bulk.  This analysis (done on the actual code)
fully explains the scalability limit displayed in
\figurename~\ref{fig:dirac-speedup}, since the overall execution time can not
be decreased when communication time becomes the limiting factor.

From this analysis for the $32^3 \times 48$ lattice, we obtain that we have a perfect
\textit{Strong Scaling} if lattice tiles associated to each GPU are at least $6$ sites thick, that is 
each GPU processes a $32^3 \times 6$ slice of the lattice.
This analysis also explains the behavior of the larger $32^3 \times 64$ lattice in 
\figurename~\ref{fig:dirac-speedup}: we should have a perfect \textit{Strong Scaling} up
to $10$ GPUs and then reach the plateau. 
As $64$ is not divisible by $10$ we cannot test precisely this configuration, but, using $16$ 
GPUs, we see a $10 \times$ speedup, corresponding to the expected plateau figure. 

\begin{figure}
\centering
\includegraphics[width=0.8\textwidth]{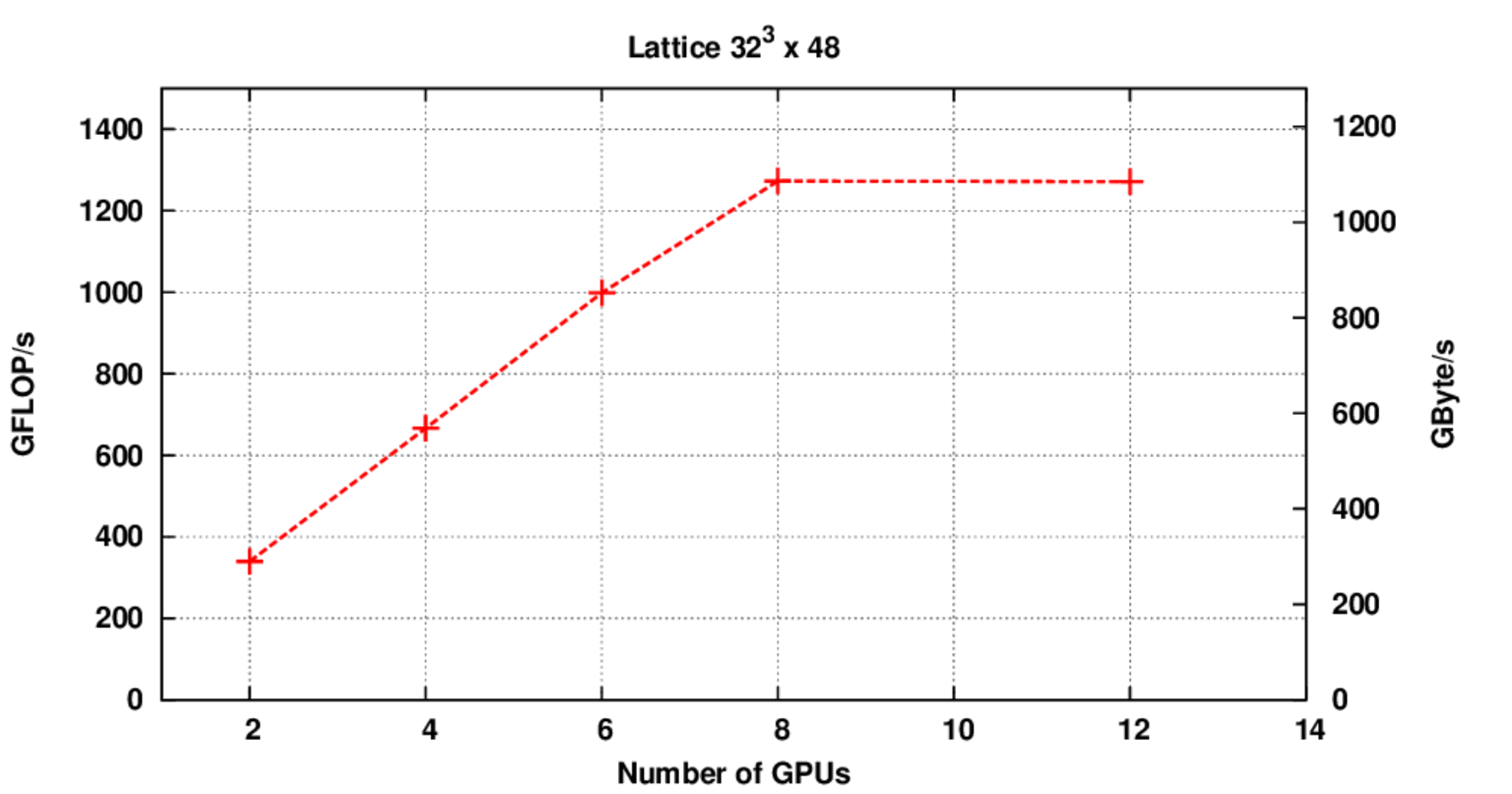}
\caption{
Aggregate performance (GFLOP/s) and bandwidth (Gbyte/s) for the Dirac operator
om a $32^3 \times 48$ lattice. Several runs of the same benchmark have been
performed with different numbers of GPUs in the COKA Cluster.}
\label{fig:dirac-perf}
\end{figure}

In order to convert our scaling results into absolute performance figures, we
have counted the floating point operations and memory accesses needed to apply
the Dirac operator to each lattice site, directly accessing GPU hardware
counters (through the PGI Profiler) and then double checking the results
against theoretical expectations.  From these measured values we have computed
the actual sustained performance (floating point operations per second) and
bandwidth (data bytes per second).  This results are shown in
\figurename~\ref{fig:dirac-perf} for several runs using an increasing number of
GPUs, on a $32^3 \times 48$ lattice. Note that we include in our operation
count the additional computational load associated to the data compression
techniques that we have described in  section \ref{sec:data-struct}.

As an example, using both GPUs of a NVIDIA K80 board, as shown in
\figurename~\ref{fig:dirac-perf}, our implementation of the staggered Dirac
Operator has a performance of $\simeq 339$ GFLOP/s in double precision and a
sustained bandwidth of $\simeq 290$ GBytes/s (partially given also by cache
accesses), that is $\approx 80\%$ of the aggregated raw peak memory bandwidth
of the processor (taking into account the bandwidth penalty associated to
Error-correcting-Codes (ECC) that we use throughout to increase data
reliability).  These figures are consistent with those obtained by other codes
adopting non-portable architecture-specific languages\cite{milc2017}.

\subsection{Gauge part of the Molecular Dynamics}
\label{sec:pg-scaling}

\begin{figure}
\centering
\includegraphics[width=0.8\textwidth]{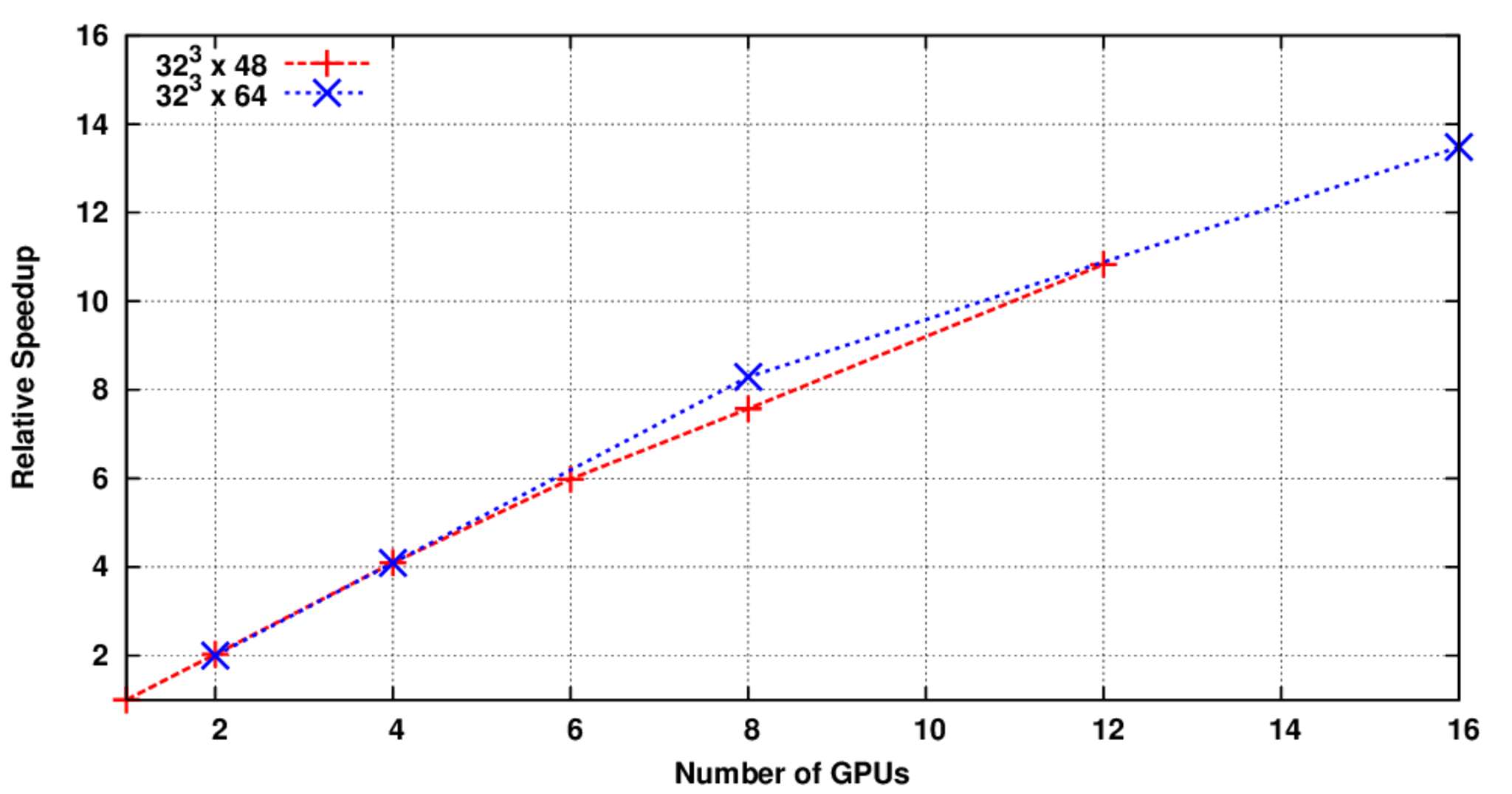}
\caption{Relative speedup of the pure gauge part of the molecular dynamics kernel 
for two different lattices of size $32^3 \times 48$ and $32^3 \times 64$, as a function of the number of GPUs.}
\label{fig:pg-speedup}
\end{figure}

We now consider the Gauge part of the molecular dynamics, which is the second
most time consuming step of the whole Monte Carlo code, after the Dirac
operator.

As shown in \figurename~\ref{fig:pg-speedup}, for this phase of the simulation
an almost perfect speedup can be appreciated up to $12$ and $16$ GPUs,
respectively for the $32^3 \times 48$ and the $32^3 \times 64$ lattices (the
same lattice sizes considered  in \figurename~\ref{fig:dirac-speedup} for the
Dirac operator).
This different behavior can be explained by the fact that the computation-time
versus communication-time ratio is more favorable in this case: to very first
approximation, for each lattice site we use a data set which is $32$ times
larger than for the Dirac operator, but the operation count increases by
$\approx 300$. As a consequence, communications do not become the scaling
limiting factor. Indeed, the true limiting factor encountered when increasing
the number of compute devices (\ie GPUs for this test), is the length of the
tiling dimension. In fact, here lattice borders have a thickness of $2$ lattice
sites, so \figurename~\ref{fig:pg-speedup} shows an almost perfect scaling up
to the point at which  the bulk size reaches zero (since $48/12 = 64/16 = 4$)
and the lattice cannot be further divided across more devices.

%%%%%%%%%%%%%%%%%%%%%%%%%%%%%%%%%%%%%%%%%%%%%%%%%%%%%%%%%%%%%%%%%%%%%%%%

\subsection{Full Simulation}

To have a comprehensive view of the performance and scaling behavior of a
complete run, as a representative example, we choose a simulation of QCD with 2
light (up and down) and 1 intermediate (strange) flavors over a $32^3 \times
48$ lattice, that is part of a production run regarding the study of QCD at
finite baryon density and towards the chiral limit.  In this particular
simulation (with quark mass $0.0015$ in lattice units and $\beta = 3.3600$),
the quark mass is about 1/3 of its physical value and the lattice spacing is
around 0.3 fermi ($3 \times 10^{-16}$ meters).  For the computations in the
molecular dynamics we are using floating-point operations, while for the
Metropolis test we are using double-precision ones.

\begin{figure}
\centering
\includegraphics[width=0.8\textwidth]{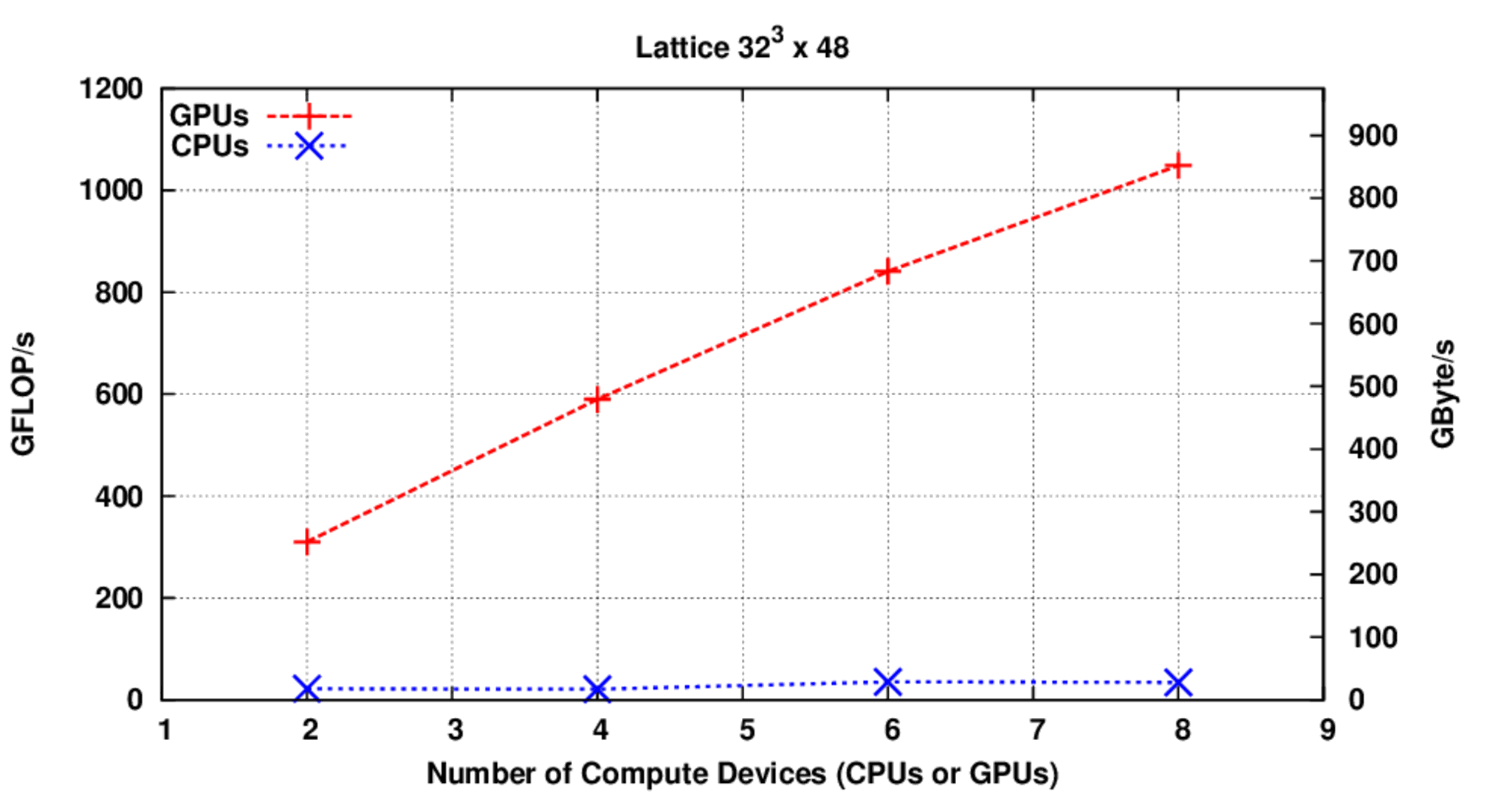}
\caption{
Aggregate performance (GFLOP/s) and bandwidth (read/write GByte/s) for a full
simulation on a $32^3 \times 48$ lattice (see the text for a list of simulation
parameters). The same simulation has been performed  using different numbers of
processors (CPUs or GPUs) in the COKA Cluster.}
\label{fig:full-sim}
\end{figure}

We have run the same simulation on different numbers of CPUs and GPUs available
on the COKA cluster, demonstrating the actual code portability offered by the
OpenACC programming model, and measuring  performance.
Our results are shown in \figurename~\ref{fig:full-sim}, as a function of the
number of computing devices.  We plot the aggregate performance (floating point
operations per second) and aggregate Memory Bandwidth (data bytes per second).
These metrics allow to appreciate both the strong scaling behavior and the
differences in absolute performance between the two architectures.

In \figurename~\ref{fig:full-sim} we see that GPUs have much higher performance
than CPUs for this kind of application.  This was partly expected, since, when
using just one processor, we had already measured a performance gap of
approximately one order of magnitude\cite{lqcdoacc17}; this gap widens further
when using multiple devices.  As an example using two CPUs. execution times
increases $\approx 14 \times$ w.r.t. two GPUs.
In order to put this figure in perspective, we note that the limitations of the
compiler when targeting CPUs, that we had described in our previous
work\cite{lqcdoacc17}, still hold today; however, when using multiple devices,
the performance measured on CPUs is further degraded by two main factors:

\begin{itemize}
\item The COKA cluster is a GPU dense machine, as each node hosts $16$ CUDA devices. This 
translates to the fact that all the GPU related points of \figurename~\ref{fig:full-sim} are 
given by simulations executing on a single-node, without requiring inter-node communications, 
but only faster intra-node data transfers between GPU memories.
On the other side, each node hosts only  two CPUs, so CPU related points in
\figurename~\ref{fig:full-sim} refer to simulations run respectively on 1, 2, 3 and 4 nodes.
\item When running on GPUs, communications can be overlapped with
communications and small kernels (such as computations on the lattice borders)
can run concurrently on the same device, as shown in
\figurename~\ref{fig:dirac-profile}.  On the other side, the current version of
the PGI compiler completely ignores -- when targeting CPUs -- \textit{async}
clauses to OpenACC directives, and fully serialize the execution of different
kernels and communications.
\end{itemize}

These factors can be, at least partially, overcome by: i) the use of CPU ``denser'' machines \eg using 
CPUs with an higher core number or even Xeon Phi processors\cite{lqcd-knl-review}; ii) the use of a 
compiler able to exploit the \textit{async} OpenACC clause.

%%%%%%%%%%%%%%%%%%%%%%%%%%%%%%%%%%%%%%%%%%%%%%%%%%%%%%%%%%%%%%%%%%%%%%%%
We also expect that single-CPU performance can be significantly increased, with
improved data layouts. Remember that the data-layout chosen for this code (\ie
Structure of Array) allows code vectorization on both GPU and CPU
architectures\cite{se4hpcs15}; however, recent
work\cite{lqcd-knl,lqcd-knl-review,dirac-opt} has shown that more complex data
structures are able to increase the performance of memory sub-system
performance on Intel CPUs and also Xeon-Phi accelerators. This has been
demonstrated also for other lattice-based simulation codes\cite{ijhpca17}.

We conclude at this stage that our code is actually portable to clusters
adopting either Intel CPU or NVIDIA GPU architectures, but an additional effort
is needed in the direction of  \emph{performance} portability, both on the
programming side and on the side of improved compiler support for CPU
architectures.  In this latter regard we add that the community developing the
GCC compiler seem to be strongly committed in supporting OpenACC and GCC,
version 7, is expected  to compile OpenACC codes for x86 CPU architecture.

\section{Conclusions}
\label{sec:conclusions}

In this work we have presented a full state-of-the-art production-grade 
code for Lattice QCD simulations with staggered fermions, coded using 
the OpenACC directive-based programming model to make the code portable 
across different computing architectures, and MPI to allow the code to run 
on multi-node systems where each node may have more then one accelerator 
installed.

This work extends the code we have developed in a previous work\cite{lqcdoacc17} 
designed to run on single-accelerator systems, turning it into a fully working MPI 
version able to exploit multi-node HPC clusters and multiple accelerators board  
within each computing node.
We have described our implementation, detailing how parallelization has been exploited, 
and the strategies adopted to improve parallel efficiency, such as overlapping of 
MPI transfers with computation to hide communications time overheads.
We have measured the scalability behavior of our code on the COKA GPU-based 
HPC cluster, and make also some tests on Intel CPU architectures.
%% Fabio
In this first implementation we have decides to use the 1d-tiling strategy, 
to keep the code structure simple, avoiding to handle non contiguous data 
communications, that for GPU-based clusters are not easily manageable\cite{parco16}, 
and also easily exploiting computation and communication overlap.
As already commented in Sec.~\ref{sec:parallelization}, this basic strategy 
keeps communication time constant while increasing the number of nodes, and for this 
reason the code scales as long as the communication time is hidden by the 
computation time.

In conclusion, our final result is a LQCD Monte Carlo code portable on a large
subset of HPC clusters, based on both GPUs and standard CPUs, with satisfactory
figures of aggregate performance and scalability. Performances measured on CPUs
are lower compared with that of GPUs; this results is inline to what we have
measured in our previous work, and the main reason is that the compiler does
not fully yet support this architecture.
We would like to highlight that this has not been a mere exercise on performance 
scalability: our efforts have been driven by the actual need to scale on multi-GPU 
architecture (basically for large RAM requirements) within the context of a project
regarding QCD at finite baryon density, for which the present code has been already 
in full production since several months.

In the near future we plan to further optimize our code for Intel processors, without 
impacting the performance on NVIDIA GPUs, hoping for a contextual further development 
of the available compilers.
We plan also to carefully assess the performance and scalability of 
our code on Intel KNL Xeon Phi clusters, as soon as the support for this 
architecture is added to the PGI compiler, or as soon as other compilers become available.
On a longer time scale, we also plan to split the lattice across more dimensions 
and investigate the impact on performance, scalability and code maintainability, 
and to investigate the impact of different memory layout to improve vectorization 
of codes especially on multi-core CPUs.

%%%%%%%%%%%%%%%%%%%%%%%%%%%%%%%%%%%%%%%%%%%%%%%%%%%%%%%%%%%%%%

\section*{Acknowledgments}

EC and FN acknowledge financial support from the INFN HPC\_HTC project.
We thank the INFN Computing Center in Pisa for providing us 
with the development framework, and Universit\`a degli Studi di Ferrara 
and INFN-Ferrara for granting access to the COKA cluster. 
This work has been developed in the framework of the COKA 
and COSA projects of INFN.

%%%%%%%%%%%%%%%%%%%%%%%%%%%%%%%%%%%%%%%%%%%%%%%%%%%%%%%%%%%%%%

%% inline BBL

%%%%%%%%%%%%%%%%%%%%%%%%%%%%%%%%%%%%%%%%%%%%%%%%%%%%%%%%%%%%%%%%%%%%%%%%

\end{document}